\documentclass[a4paper, 11 pt]{JHEP3}
\usepackage{epsfig,bbm}

\usepackage{ifthen}
\usepackage{epsfig}
\usepackage{amssymb}
\usepackage{graphicx}
\usepackage{amsmath}
\usepackage[center]{subfigure}
\renewcommand{\text}[1]{%
\ifthenelse{\equal{#1}{mB}}{m_B}{}%
\ifthenelse{\equal{#1}{sig}}{\sigma}{}%
\ifthenelse{\equal{#1}{mc}}{m_c}{}%
\ifthenelse{\equal{#1}{q2}}{q^2}{}%
\ifthenelse{\equal{#1}{om}}{\omega}{}%
}
\newcommand{\ba}{\begin{eqnarray}}
\newcommand{\ea}{\end{eqnarray}}
\newcommand{\be}{\begin{equation}}
\newcommand{\ee}{\end{equation}}

\title{Heavy-to-light baryonic form factors at large recoil}

\author{Thomas Mannel and  Yu-Ming Wang \\
Theoretische Elementarteilchenphysik, \\ Naturwissenschaftlich Techn. Fakult\"at \\
Universi\"at Siegen, 57068 Siegen, \\ Germany}

\abstract{ We analyze  heavy-to-light baryonic form factors at
large recoil and derive the scaling behavior of these form factors
in the heavy quark limit. It is shown that only one universal form
factor is needed to parameterize $\Lambda_b \to p$ and $\Lambda_b
\to \Lambda$ matrix elements in the large recoil limit of light
baryons, while hadronic matrix elements of $\Lambda_b \to \Sigma$
transition vanish in the large energy limit  of $\Sigma$ baryon
due to the space-time parity symmetry. The scaling law of the soft
form factor $\eta(P^{\prime} \cdot v)$, $P^{\prime}$ and $v$ being
the momentum of nucleon and the velocity of $\Lambda_b$ baryon,
responsible for $\Lambda_b \to p$ transitions is also derived
using the nucleon distribution amplitudes in  leading conformal
spin. In particular, we verify that this scaling behavior is in
full agreement with that from light-cone sum rule approach in the
heavy-quark limit. With these form factors, we further investigate
the $\Lambda$ baryon polarization asymmetry $\alpha$ in $\Lambda_b
\to \Lambda \gamma$ and the forward-backward asymmetry $A_{FB}$ in
$\Lambda_b \to \Lambda l^{+} l^{-}$.  Both two observables
($\alpha$ and $A_{FB}$) are independent of hadronic form factors
in  leading power of $1/m_b$ and in  leading order of $\alpha_s$.
We  also extend the analysis of hadronic matrix elements for
$\Omega_b \to \Omega$ transitions to rare $\Omega_b \to \Omega \,
\gamma$ and $\Omega_b \to \Omega \, l^{+} l^{-}$ decays and find
that radiative $\Omega_b \to \Omega \, \gamma$ decay is probably
the most promising FCNC $ b \to s$ radiative baryonic decay
channel. In addition, it is interesting to notice that the
zero-point of forward-backward asymmetry of $\Omega_b \to \Omega
\, l^{+} l^{-}$ is the same as the one for $\Lambda_b \to \Lambda
l^{+} l^{-}$ to leading order accuracy provided that the form
factors $\bar{\zeta}_i$ ($i=3, \, 4, \, 5\,$) are numerically as
small as indicated from  the quark model. }


\begin{document}

\section{Introduction}

Heavy baryons have attracted renewed attention due to the
expectation of the future data on processes involving these
states. In particular, LHCb will open a window on these particles
by producing a sizable number of bottom and also charmed baryons.

From the theoretical side these states have been investigated
already in the early days of Heavy Quark Effective Theory (HQET)
\cite{Isgur:1990pm,Mannel:1990vg,Falk:1993rf}. In fact, from the
point of view of HQET  the $\Lambda_b$ baryon is the simplest
state,  since the light degrees of freedom are in a spin and
isospin singlet state, hence the $\Lambda_b$ spin is equal to the
$b$ quark spin.

However, HQET is applicable only in cases where the light degrees
of freedom do not carry a large momentum in  the rest frame of the
heavy hadron. Thus the kinematics of processes which can be
described in this way is strongly restricted. For semileptonic
decays this means that the methods can be applied only for
leptonic momentum transfer $q^2$ close to $q^2_{\rm max}$, where
the recoil on the light degrees of freedom is small.

Various attempts have been made to formulate an effective theory
for the situation where a weak process generates energetic light
degrees of freedom in the rest frame of a decaying heavy hadron.
For semileptonic processes this is the region close to $q^2 = 0$,
where the light quark has an energy of the order of the mass of
the heavy hadron. The first attempt, called Large Energy Effective
Theory (LEET) \cite{Dugan:1990de,Charles:1998dr}, had certain
problems which were eventually cured by Soft Collinear Effective
Theory (SCET) \cite{Bauer:2000yr,Beneke:2002ph}, which is now
considered to be the appropriate description of energetic light
quarks and gluons.

Both LEET as well as SCET exhibit additional symmetries which are
equivalent to the conformal spin symmetry of massless QCD. These
symmetries have been used to restrict the number of form factors
in the semileptonic decays of a $B$ meson into a pion and a $\rho$
meson to only three independent unknown functions
\cite{Charles:1998dr}. SCET allows to compute corrections to these
relations which hold in the infinite energy limit of the outgoing
light meson \cite{Beneke:2002ph,Beneke:2003pa}.

The purpose of the present paper is to apply the same methods to
the case of baryonic transitions. It tuns out that (similar to the
HQET application for soft light degrees of freedom) a significant
reduction of the number of form factors is achieved also for the
case of energetic light degrees of freedom. It turns out that in
the infinite energy limit only a single form factor is needed to
describe $\Lambda_b \to p$ and $\Lambda_b \to \Lambda$
transitions, while the $\Lambda_b \to \Sigma$ transition vanishes
in this limit. The analysis is also extended to $\Omega_b \to
\Omega$ transitions, which may also have interesting
phenomenological applications.

In the next sections we shall briefly review the necessary
ingredients of HQET and LEET/SCET. Section 3 is devoted to  the
derivation of the necessary tensor representations of heavy
baryons and the light-cone projectors of light baryons consisting
of energetic collinear quarks.  The core relations for weak form
factors are derived in section 4, while section 5 contains the
phenomenological applications  to $\Lambda_b$ and $\Omega_b$
decays, where we focus on radiative and semileptonic FCNC decays,
which may be an interesting target at LHCb. The concluding
discussion is presented in section 6.

\section{Brief Review of HQET and SCET}


HQET is constructed using the limit $m_Q \to \infty$, where $m_Q$ is the mass
of the heavy quark.
In this limit, the dynamics of heavy-light hadron
system can be greatly simplified due to new symmetries which are
absent in full QCD.  The heavy quark in the heavy hadron
acts as a static color source which binds the light degrees of freedom
by the exchange of soft gluons. Since the color interaction of QCD
is flavor blind, the light degrees of freedom are in the same state independent
of the flavour of the heavy quark. Likewise, since the chromomagnetic moment
of the heavy quark scales with $1/m_Q$, the spin of the heavy quark decouples
in the infinite mass limit.

The momentum of the heavy quark $p_Q$ bound in a heavy hadron
moving with the four-velocity is decomposed according to
\begin{eqnarray}
p_Q^{\mu}=m_{Q} v^{\mu} + k^{\mu}\,,
\end{eqnarray}
where the residual momentum is small,
$k_{\mu} \sim {\rm \Lambda}_{QCD}$ and reflects the off-shell
fluctuations  due to the soft interactions.

The heavy quark field $Q(x)$ can be decomposed using the velocoy vector $v$ as
\begin{eqnarray}
h_v(x)&=&{1+ \not \! v \over  2} {\rm exp } \{i m_Q v \cdot x\}
Q(x) \,, \nonumber \\
H_v(x)&=&{1 - \not \! v \over  2} {\rm exp } \{i m_Q v \cdot x\}
Q(x) \,,
\end{eqnarray}
which  correspond to the large and small components of $Q(x)$.

The field $H(x)$ corresponds to massive fluctuations related to
the scale $2 m_Q$, while the field $h(x)$ corresponds to a
massless excitation. Integrating out the the massive degree of
freedom, we end up with the effective Lagrangian of HQET
%
%
%
\cite{Isgur:1989vq,Isgur:1989ed,Eichten:1989zv,Grinstein:1990mj,Georgi:1990um}
\begin{eqnarray}
\mathcal{L}_{\rm HQET}=\bar{h}_v (i v \cdot D) h_v + O(1/m_{Q})
\,,
\end{eqnarray}
 where
$D_{\mu}=\partial_{\mu} - ig A_{\mu}$ denotes the covariant derivative of QCD, involving the
gluon field.

Note that the leading order interaction is  independent on the
heavy-quark mass and shows explicitly the heavy-flavor and spin
symmetry mentioned above. Formally, the  HQET Lagrangian is
invariant under the transformation of ${\rm SU}(n_{Q})_{flavor}
\otimes {\rm SU}(2)_{spin}$ group, where $n_Q$ denotes the number
of heavy quarks \footnote{In fact, the symmetry is even larger, it
is actually ${\rm SU}(2 n_Q)$.}.

Energetic light degrees of freedom are described in SCET, which
was predated by an attempt to formulate an effective theory
involving energetic partons (LEET) \cite{Dugan:1990de}.
%
However, it was soon observed
\cite{Aglietti:1997zk,Balzereit:1998yf} that LEET could not
reproduce the the infrared physics of full QCD due to the absence
of  collinear gluon interaction with energetic quarks. A nonlocal
counter term was introduced in \cite{Balzereit:1998yf} to cancel
emerged non-local divergence, however large logarithms still
remains in the matching coefficient from QCD to LEET. This development
finally lead to the formulation of SCET.  In Ref.
\cite{Bauer:2000yr}, SCET involving both soft and collinear gluons
coupling to energetic partons was formulated and infrared physics
of QCD can be reproduced correctly. However, the inclusion of
collinear models does not change the relations of soft form
factors in the large energy limit. 

Massless QCD as well as SCET has, to leading order, a conformal
spin symmetry which - similar to the symmetries of HQET - lead to
relations among form factors in the symmetry limit. In particular,
applying this to heavy-to-light transition \cite{Charles:1998dr}
one obtains relations among heavy-to-light
mesonic form factors. 

Following closely to the derivation of HQET, the momentum of the
energetic\footnote{This of course implies the definition of a
reference frame in which the parton is energetic. For our purposes
this frame is defined to be the rest frame of the decaying heavy
hadron.} parton can be split  as
\begin{eqnarray}
p_{q}^{\mu}= E n^{\mu} + \tilde{k}^{\mu}, \hspace{1 cm}
|\tilde{k}| \ll E
\end{eqnarray}
where $E$ is the energy of light hadron and $n = p/E$ is the
light-like vector in the direction of the outgoing light decay
products. Making use of the velocity $v$ of the decaying heavy
hadron one may define a second light-like vector such that
$$
v = \frac{1}{\sqrt{2}} (n+\bar{n}) \, , \quad n^2 = 0 = \bar{n}^2
\, , \quad n \bar{n} = 1 \, .
$$
The quark field of full QCD
$q(x)$ can be decomposed according to
\begin{eqnarray}
q_n(x)&=& {\not \! n  \not \! \bar{n} \over  2} {\rm exp } \{i E
n \cdot x\} q(x) \,, \nonumber \\
q_{\bar{n}}(x)&=& {\not \! \bar{n}  \not \! n \over  2} {\rm exp }
\{i E n \cdot x\} q(x) \,,
\end{eqnarray}
where the field $q_{\bar{n}}(x)$ is related to the large energy scale $2 E$, while the field
$q_n(x)$ contains the small energy scales. Similar to the case of HQET we may integrate out the
field $q_{\bar{n}}(x)$ and obtain the leading order SCET Lagrangian
%
\begin{eqnarray}
\mathcal{L}_{\rm SCET}=\bar{q}_n \left[ i n \cdot D + i \not \!
D_c^\perp \frac{1}{2 \, i \, \bar{n} \cdot D_c} i \not \!
D_c^\perp \right] {i \not \! \bar{n}} \, q_n + O(1/E)\,,
\label{effective lagrangians: LEET}
\end{eqnarray}
where the soft gluon field has been separated out in
$D_c^{\mu}=\partial^{\mu}-i g A_{c}^{\mu}$ with $A_{c}$ being the
collinear component of the gluon field. It is worthwhile to point
out that the projector prosperities of the collinear fields remain
in this leading order Lagrangian \cite{Beneke:2002ph}, resulting
in the symmetry relations between form factors observed in
\cite{Charles:1998dr}.


\section{Tensor Representations and Light-cone projectors of baryons}

In the following we discuss the transition matrix elements of
heavy-to-light transitions for  baryons at large recoil, i.e. the
outgoing light baryon carries a large energy in the rest frame of
the decaying heavy baryon. Thus there are various degrees of
freedom involved: the heavy quark in the initial state accompanied
by light degrees of freedom which are soft, with   momenta of
 order  $\Lambda_{\rm QCD}$; furthermore, the final state baryon with a
large energy is assumed to  consist of  three collinear quarks.
Since we shall exploit the collinear spin symmetry, we will make
the spin indices of these collinear quarks explicitly by using the
light-cone projectors  of the light, energetic baryon states. Weak
transition matrix elements can then be considered by a
generalization of the well known trace formulae used in the case
of the  mesonic transitions \cite{Falk:1990yz}.


\subsection{Tensor representations for heavy baryons}

Following  Refs. \cite{Isgur:1990pm,Mannel:1990vg}, the
nomenclature of low-lying heavy baryons is
\begin{eqnarray}
\Lambda_Q = [(q q^{\prime})_0  \, Q]_{1/2}\,,  & \qquad & \Xi_Q =
[(q
s)_0 \, Q]_{1/2}\,, \nonumber  \\
\Sigma_Q = [(q q^{\prime})_1  \, Q]_{1/2}\,,  & \qquad &
\Xi^{\prime}_Q = [(q s)_1 \, Q]_{1/2}\,, \nonumber  \\
\Omega_Q = [(s s)_1  \, Q]_{1/2}\,,  & \qquad &
\Sigma^{\ast}_Q = [(q q^{\prime})_1 \, Q]_{3/2}\,, \nonumber  \\
\Xi_Q^{\ast} = [(q s)_1  \, Q]_{3/2}\,,  & \qquad &
\Omega^{\ast}_Q = [(s s)_1 \, Q]_{3/2}\,.
\end{eqnarray}
The heavy baryon states $\Lambda_Q$ and $\Xi_Q $ are exactly
analogous and identical in the SU(3) flavor symmetry limit. The
other baryons shown in the above correspond to the heavy-baryon
sextet categorized by the spin-parity of light-quark system,
$(\Sigma_Q, \Sigma^{\ast}_Q)$, $(\Xi^{\prime}_Q, \Xi_Q^{\ast})$
and $(\Omega_Q, \Omega^{\ast}_Q)$  degenerate in the heavy quark
limit. For this reason, we will concentrate on the tensor
representations of $\Lambda_Q$ and $\Omega_Q$ baryons below.

The representation of $\Lambda_b$ baryon is trivial since the soft
light degrees of freedom are in a spinless state and  the bottom
quark carries all of the angular momentum of the baryon in the
heavy quark limit. Thus we have
\begin{eqnarray}
 | \Lambda_b \rangle \mapsto   \Lambda_b(v)\equiv b(v) \,,
\end{eqnarray}
where $\Lambda_b(v)$ ($b(v)$) on the right hand just denotes the
Dirac spinor of $\Lambda_b$ baryon ($b$-quark) with velocity $v$.

The representation of  $\Omega_Q^{(\ast)}$ baryons involve some
nontrivial Lorenz structures, since the light degrees of freedom
now carry spin.  Two equivalent representations  (pseudovector and
antisymmetric tensor) can be introduced  to describe these states
\cite{Mannel:1990vg}. Here, we will stick to the pseudovector
representation
\begin{eqnarray}
 | \Omega_Q^{(\ast)} \rangle \mapsto   R^{\mu}(v) = A^{\mu}  b(v)\,,
\end{eqnarray}
following the Ref. \cite{Falk:1991nq}, where $R^{\mu}(v)$
satisfies the constraint of transversality.

The  pseudovector-spinor object $R^{\mu}$ can be decomposed as
\begin{eqnarray}
R^{\mu}_{3/2}=[g_{\mu \nu} - {1 \over 3} (\gamma^{\mu}+v^{\mu})
\gamma_{\nu}] R^{\nu}
\end{eqnarray}
for a spin-${3 \over 2}$ baryon and
\begin{eqnarray}
R^{\mu}_{1/2}={1 \over 3} (\gamma^{\mu}+v^{\mu}) \gamma_{\nu}
R^{\nu}={1 \over \sqrt{3}} (\gamma^{\mu}+v^{\mu}) \gamma_5 h_v
\end{eqnarray}
for a spin-${1 \over 2}$ baryon.  $R^{\mu}_{3/2}$  satisfies the
properties of Rartia-Schwinger vector spinor
\begin{eqnarray}
\not \! v  R^{\mu}_{3/2} = R^{\mu}_{3/2}\,,  \hspace{2 cm} v_{\mu}
R^{\mu}_{3/2}=0 \,, \hspace{2 cm} \gamma_{\mu}R^{\mu}_{3/2}=0 \,;
\end{eqnarray}
and $h_v={1 \over \sqrt{3}} \gamma_5 \gamma_{\mu} R^{\mu}_{1/2}$
is just the Dirac spinor of $\Omega_b({1 \over 2}^{+})$ state
moving with the velocity $v_{\mu}$.

\subsection{Light-cone projectors for  light baryons}

We assume that the quarks in the nucleon are collinear quarks, and
hence we have collinear spin symmetries for these quarks. To the
leading twist accuracy,  the nucleon DA's at $z^2\to 0$ are
defined according to \cite{Braun:2000kw}:
\begin{eqnarray}
\langle 0 | \epsilon^{ijk} u_\alpha^i(a_1 z) u_\beta^j(a_2 z )
d_\gamma^k(a_3 z) | N(P) \rangle &=& \mathcal{V}_N
\left(\!\not\!{P}C \right)_{\alpha \beta} \left(\gamma_5 N
\right)_\gamma  + \mathcal{A}_N \left(\!\not\!{P} \gamma_5 C
\right)_{\alpha \beta}
\left(N \right)_\gamma    \nonumber \\
&& + \mathcal{T}_N\left(P^\nu i \sigma_{\mu\nu} C\right)_{\alpha
\beta} \left(\gamma^\mu\gamma_5 N \right)_\gamma .
\label{nucleon:DA}
\end{eqnarray}
where $\alpha,\beta,\gamma$ are Dirac indices and $z$ is a
light-ray vector $z^2=0$. The calligraphic notations
$\mathcal{F}=\{\mathcal{V}_N, \mathcal{A}_N,\mathcal{T}_N\}$
denote the integrals over the twist-3 nucleon DA's:
\begin{eqnarray}
\mathcal{F}=\int \mathcal{D} x \,\,  {\rm exp } \bigg[-i \sum
\limits_{i=1}^{3} a_i x_i P \cdot z \bigg] F(x_i,\mu)\,,
\label{eq:F}
\end{eqnarray}
represented by the same non-calligraphic  letters
$F=\{V_N,A_N,T_N\}$.  Here,  $x_i=\{x_1,x_2,x_3\}$ with $0 \leq
x_i \leq 1$ are  the longitudinal momentum fractions of the quarks
in the nucleon,  $\mu$ is the normalization scale and the integral
measures read
\begin{eqnarray}
\int \mathcal{D} x \equiv \int d x_1 d x_2 d x_3
 \delta(1-x_1-x_2-x_3) \,.
\end{eqnarray}
This definition is equivalent to the following structure of the
nucleon state \cite{Chernyak:1987nu,Farrar:1988vz}
\begin{eqnarray}
|N^{\uparrow} (P) \rangle = f_{N} \int {\mathcal{D} x \over 4
\sqrt{24 x_1 x_2 x_3} }  \bigg \{ V_N(x_i) \big | \big(
u^{\uparrow}(x_1) u^{\downarrow}(x_2) + u^{\downarrow}(x_1)
u^{\uparrow}(x_2) \big) d^{\uparrow}(x_3) \big \rangle &&
\nonumber
\\
 - A_N(x_i) \big | \big( u^{\uparrow}(x_1) u^{\downarrow}(x_2) -
u^{\downarrow}(x_1) u^{\uparrow}(x_2) \big) d^{\downarrow}(x_3)
\big \rangle  && \nonumber
\\
- 2 T_N(x_i) \big |  u^{\uparrow}(x_1) u^{\uparrow}(x_2)
d^{\downarrow}(x_3) \big \rangle\bigg \}. && \label{necleon
configuration}
\end{eqnarray}
The spin symmetry of $[u\,u]$ diquark implies that the
distribution amplitudes $V_{N}$, $A_{N}$ and $T_{N}$ satisfy the
following relations
\begin{eqnarray}
V_{N}(x_1, x_2, x_3)&=&V_{N}(x_2, x_1, x_3),\, \qquad A_{N}(x_1,
x_2, x_3)=-A_{N}(x_2, x_1, x_3),\, \nonumber
\\ T_{N}(x_1, x_2, x_3)&=&T_{N}(x_2, x_1, x_3),
\end{eqnarray}
hence the distribution amplitude $A_1(x_i)$ vanishes in the
leading conformal spin approximation, and we will drop out this
term in the following analysis.

It is straightforward to derive the tensor representation of the
nucleon state in the collinear limit, namely the nucleon
light-cone projectors
\begin{eqnarray}
\mathcal{M}^{V}_{N} &=& (\not \! n    C )_{\alpha
\beta} \,\,  [\gamma_5 N]_{\gamma} \,, \nonumber \\
\mathcal{M}^{T}_{N} &=& (n^{\nu} i \sigma_{\mu \nu}   C )_{\alpha
\beta} \,\, [\gamma^{\mu} \gamma_5 N]_{\gamma}  \,,
\end{eqnarray}
corresponding to the first and third configurations in  Eq.
(\ref{necleon configuration}), where the momentum of nucleon
$P_{\mu}$ is chosen along the light-ray $n_{\mu}$ direction and
$\xi_n$ is the Dirac spinor of a fermion moving on the collinear
direction $n_{\mu}$ ($\not \! n \xi_n =0$). Likewise, the
light-cone projectors for the nucleon in the final state  can be
written as
\begin{eqnarray}
\mathcal{\overline{M}}^{V}_{N} &=& -(C \not \! n  )_{\alpha
\beta} \,\,  [\bar{N} \gamma_5 ]_{\gamma} \,, \nonumber \\
\mathcal{\overline{M}}^{T}_{N} &=& -(n^{\nu} i C  \sigma_{\mu
\nu})_{\alpha \beta} \,\, [\bar{N} \gamma^{\mu} \gamma_5
]_{\gamma} \,.
\end{eqnarray}

Similarly, one can define the leading-twist distribution
amplitudes of $\Lambda$ baryon \cite{Liu:2008yg}
\begin{eqnarray}
\langle 0 | \epsilon^{ijk} u_\alpha^i(a_1 z) d_\beta^j(a_2 z )
s_\gamma^k(a_3 z) | N(P) \rangle &=& \mathcal{V}_{\Lambda}
\left(\!\not\!{P}C \right)_{\alpha \beta} \left(\gamma_5 \Lambda
\right)_\gamma  + \mathcal{A}_{\Lambda} \left(\!\not\!{P} \gamma_5
C \right)_{\alpha \beta}
\left(\Lambda \right)_\gamma    \nonumber \\
&& + \mathcal{T}_{\Lambda}\left(P^\nu i \sigma_{\mu\nu}
C\right)_{\alpha \beta} \left(\gamma^\mu\gamma_5 \Lambda
\right)_\gamma \,, \label{Lambda:DA}
\end{eqnarray}
which is equivalent to the following structure of the $\Lambda$
baryon state
\begin{eqnarray}
|\Lambda^{\uparrow} (P) \rangle =  \int {\mathcal{D} x \over 4
\sqrt{24 x_1 x_2 x_3} }  \bigg \{f^V_{\Lambda} V_{\Lambda}(x_i)
\big | \big( u^{\uparrow}(x_1) d^{\downarrow}(x_2) +
u^{\downarrow}(x_1) d^{\uparrow}(x_2) \big) s^{\uparrow}(x_3) \big
\rangle && \nonumber
\\
 - f^V_{\Lambda} A_{\Lambda}(x_i) \big | \big( u^{\uparrow}(x_1) d^{\downarrow}(x_2) -
u^{\downarrow}(x_1) d^{\uparrow}(x_2) \big) s^{\downarrow}(x_3)
\big \rangle  && \nonumber
\\
- 2 f^T_{\Lambda} T_{\Lambda}(x_i) \big |  u^{\uparrow}(x_1)
d^{\uparrow}(x_2) s^{\downarrow}(x_3) \big \rangle\bigg \}. &&
\label{Lambda configuration}
\end{eqnarray}
The isospin symmetry of $[u\,d]$ diquark implies that distribution
amplitudes $V_{\Lambda}$, $A_{\Lambda}$ and $T_{\Lambda}$ respect
the following relations
\begin{eqnarray}
V_{\Lambda}(x_1, x_2, x_3)&=&-V_{\Lambda}(x_2, x_1, x_3),\, \qquad
A_{\Lambda}(x_1, x_2, x_3)=A_{\Lambda}(x_2, x_1, x_3),\, \nonumber
\\ T_{\Lambda}(x_1, x_2, x_3)&=&-T_{\Lambda}(x_2, x_1, x_3),
\end{eqnarray}
therefore $V_{\Lambda}(x_1, x_2, x_3)$ and $T_{\Lambda}(x_1, x_2,
x_3)$ vanish to the leading conformal spin accuracy, and only the
second term in Eq. (\ref{Lambda configuration}) will be considered
below. The light-cone projector of $\Lambda$ baryon can be written
as
\begin{eqnarray}
\mathcal{M}^{A}_{\Lambda} = (\not \! n  \gamma_5  C )_{\alpha
\beta} \,\,  ( \Lambda )_{\gamma} \,, \qquad
\mathcal{\overline{M}}^{A}_{\Lambda} = (C \not \! n  \gamma_5
)_{\alpha \beta} \,\,  ( \bar{\Lambda} )_{\gamma} \,,
\end{eqnarray}
for the initial  and  final states, respectively.

In addition,  the light-cone projectors of $\Sigma$ and $\Xi$
baryons are the same as the ones for the  nucleon state. Below, we
explicitly present the projectors of $\Sigma$  baryon
\begin{eqnarray}
\mathcal{M}^{V}_{\Sigma} = (\not \! n    C )_{\alpha \beta} \,\,
[\gamma_5 \Sigma]_{\gamma} \,, & \qquad & \mathcal{M}^{T}_{\Sigma}
= (n^{\nu} i \sigma_{\mu \nu}   C )_{\alpha \beta} \,\,
[\gamma^{\mu} \gamma_5 \Sigma]_{\gamma}   \,, \nonumber \\
\mathcal{\overline{M}}^{V}_{\Sigma} = -(C \not \! n )_{\alpha
\beta} \,\,  [\bar{\Sigma} \gamma_5 ]_{\gamma} \,,  & \qquad &
\mathcal{\overline{M}}^{T}_{\Sigma} = -(n^{\nu} i C \sigma_{\mu
\nu})_{\alpha \beta} \,\, [\bar{\Sigma} \gamma^{\mu} \gamma_5
]_{\gamma} \,,
\end{eqnarray}
for the initial and final states.

\subsection{Light-cone projectors of baryon decuplet}

Following Refs. \cite{Chernyak:1987nu,Braun:1999te}, the
distribution amplitudes of spin-${3 \over 2}$ $\Omega$ baryon are
defined as
\begin{eqnarray}
&& \langle 0 | \epsilon^{ijk} s_\alpha^i(a_1 z) s_\beta^j(a_2 z )
s_\gamma^k(a_3 z) | \Omega(P) \rangle \nonumber \\
&& = { \lambda_{\Omega}^{1/2} \over 4} \bigg [\mathcal{V}_{\Omega}
\left( \gamma_{\mu} C \right)_{\alpha \beta} \left(\Omega^{\mu}
\right)_\gamma  + \mathcal{A}_{\Omega} \left(\gamma_{\mu} \gamma_5
C \right)_{\alpha \beta} \left( \gamma_5 \Omega^{\mu}
\right)_\gamma - {\mathcal{T}_{\Omega} \over 2} \left( i
\sigma_{\mu\nu} C\right)_{\alpha \beta} \left(\gamma^\mu
\Omega^{\nu} \right)_\gamma \bigg ]\nonumber \\
&&  \hspace{0.5 cm} -{1 \over 4} f_{\Omega}^{3/2}\Phi_{\Omega} (i
\sigma_{\mu \nu} C)_{\alpha \beta} (P_{\mu} \Omega^{\nu}-{1 \over
2} m_{\Omega} \gamma_{\mu} \Omega^{\nu})_{\gamma}\,,
\label{Omega:DA}
\end{eqnarray}
where $f_{\Omega}^{3/2}=\sqrt{2 \over 3}
\lambda_{\Omega}^{1/2}/m_{\Omega}$ and $\Omega_{\gamma}^{\mu}$ is
the $\Omega$ resonance spin-${3 \over 2}$ vector
\begin{eqnarray}
( \not \! P -m_{\Omega} ) \Omega^{\mu} = 0 \,, \hspace{2 cm}
\bar{\Omega}^{\mu} \Omega_{\mu}=-2 m_{\Omega}\,, \hspace{2 cm}
\gamma_{\mu} \Omega^{\mu} = P_{\mu} \Omega^{\mu}=0\,.
\end{eqnarray}
The following symmetry relations among the distribution amplitudes
\begin{eqnarray}
V_{\Omega}(x_1, x_2, x_3)=V_{\Omega}(x_2, x_1, x_3)\,, & \qquad &
A_{\Omega}(x_1, x_2, x_3)=-A_{\Omega}(x_2, x_1, x_3)\,, \nonumber
\\
T_{\Omega}(x_1, x_2, x_3)=T_{\Omega}(x_2, x_1, x_3)\,, & \qquad &
\Phi_{\Omega}(x_1, x_2, x_3)=\Phi_{\Omega}(x_2, x_1, x_3)\,,
\end{eqnarray}
can be identified. The light-cone projectors of $\Omega$-baryon
can be written as
\begin{eqnarray}
\mathcal{M}^{V}_{\Omega} &=& (\gamma_{\mu}    C )_{\alpha \beta}
\,\,  [\Omega^{\mu}]_{\gamma} \,, \mathcal  \qquad
{M}^{T}_{\Omega} = ( i \sigma_{\mu \nu}   C )_{\alpha
\beta} \,\, [\gamma^{\mu} \Omega^{\nu}]_{\gamma}  \,, \nonumber \\
\mathcal{M}^{\Phi}_{\Omega} &=& ( i \sigma_{\mu \nu}   C )_{\alpha
\beta} \,\, [n^{\mu} \Omega^{\nu}]_{\gamma}  \,,
\end{eqnarray}
for the initial state and
\begin{eqnarray}
\mathcal{\overline{M}}^{V}_{\Omega} &=& (C \gamma_{\mu} )_{\alpha
\beta} \,\,  [\bar{\Omega}^{\mu}]_{\gamma} \,, \qquad
\mathcal{\overline{M}}^{T}_{\Omega} = (  C i \sigma_{\mu \nu}
)_{\alpha
\beta} \,\, [\bar{\Omega}^{\nu} \gamma^{\mu}]_{\gamma}  \,, \nonumber \\
\mathcal{\overline{}M}^{\Phi}_{\Omega} &=& (C i \sigma_{\mu \nu}
)_{\alpha \beta} \,\, [\bar{\Omega}^{\nu} n^{\mu} ]_{\gamma}  \,,
\end{eqnarray}
for the final state.

\section{Weak form factors for bottom baryon decays}

\subsection{$\Lambda_b \to p$,  $\Lambda_b \to \Lambda$ and $\Lambda_b \to \Sigma$ form factors}

Neglecting the hard interactions, $\Lambda_b \to p$ form factors
at large recoil can can be written down in terms of the tensor
representations of the participating baryons. Heavy quark and
collinear spin symmetries imply
\begin{eqnarray}
\langle N(P^{\prime}) | \bar{u} \Gamma b| \Lambda_b(v) \rangle &=&
\sum_{l=V,T} \, [\mathcal{\overline{M}}^{l}_{N}]_{\alpha
\beta,\gamma}\,
[\mathcal{J}_l C^{T}]_{\gamma \beta} \, \delta_{\alpha \tau} \,  [\Gamma \, \Lambda_b(v)]_{\tau}\nonumber \\
&=& \bar{N}(P^{\prime}) \gamma_5 \mathcal{J}_1 \not \! n \Gamma
\Lambda_b(v)  + \bar{N}(P^{\prime}) \gamma^{\mu} \gamma_5
\mathcal{J}_2  i n^{\nu} \sigma_{\mu \nu} \Gamma \Lambda_b(v)\,,
\label{Lambdab to proton}
\end{eqnarray}
where the non-perturbative dynamics is embedded in the coefficient
functions $\mathcal{J}_i$ ($i=1,2$). The most general structures
of $\mathcal{J}_i$ can be written as
\begin{eqnarray}
\mathcal{J}_i= (a_i + b_i \not \! {n} + c_i \not \! {\bar{n}} +
d_i \not \! {n}  \not \! {\bar{n}}) \gamma_5 \,,
\end{eqnarray}
where the coefficients $a_i$,  $b_i$, $c_i$ and $d_i$ are
functions of $P^{\prime} \cdot v$. Using the equation of motion
$\not \! n \xi_n =0$,  the matrix element (\ref{Lambdab to
proton}) can be reduced to
\begin{eqnarray}
\langle P(P^{\prime}) | \bar{u} \Gamma b| \Lambda_b(v) \rangle =
\bar{N}(P^{\prime}) (\eta_1 + \eta_2 \not \! v )\Gamma
\Lambda_b(v)\,, \label{Lambdab to proton: LEET}
\end{eqnarray}
which is the same as that obtained at small recoil, involving
only soft light degrees of freedom. However, we may further
simplify it by considering the  matrix element with $\Gamma=\not
\! n $, for which we have
\begin{eqnarray}
0=\langle P(P^{\prime}) | \bar{u} \not \! n  \,  b| \Lambda_b(v)
\rangle = \bar{N}(P^{\prime}) (\eta_1 + \eta_2 \not \! v )  \not
\! n \Lambda_b(v)=  \sqrt{2} \, \eta_2 \, \bar{N}(P^{\prime})
 \Lambda_b(v) \,,
\end{eqnarray}
indicating that the soft form factor $\eta_2$ vanishes at large
recoil to the leading-power accuracy.  Then,   the $\Lambda_b \to
p$ transition matrix element is written as
\begin{eqnarray}
\langle P(P^{\prime}) | \bar{u} \Gamma b| \Lambda_b(v) \rangle
=\eta(P^{\prime} \cdot v) \bar{N}(P^{\prime})  \Gamma
\Lambda_b(v)\,, \label{Lambdab to proton: EFT}
\end{eqnarray}
where only one universal form factor $\eta(P^{\prime} \cdot v)$
appears to  leading order. In  full QCD, the weak form factors of
$\Lambda_b \to p$ induced by $V-A$ current are defined as
\begin{eqnarray}
\langle N(P^{\prime}) | \bar{c} \,\gamma_\mu\, u | \Lambda_b(P)
\rangle &=& \bar{N} (P^{\prime})  \bigg\{f_1(q^2)\,\gamma_\mu +
i\frac{f_2(q^2)}{m_{\Lambda_b}}\,\sigma_{\mu\nu}q^\nu
+\frac{f_3(q^2)}{m_{\Lambda_b}}\,q_\mu\bigg\} \Lambda_b(P) \,,
\nonumber \\
\langle N(P^{\prime}) | \bar{c} \,\gamma_\mu \gamma_5 \, u |
\Lambda_b(P) \rangle &=& \bar{N} (P^{\prime})
\bigg\{g_1(q^2)\,\gamma_\mu + i\frac{g_2(q^2)}{m_{\Lambda_b}}
\,\sigma_{\mu\nu}q^\nu
+\frac{g_3(q^2)}{m_{\Lambda_b}}\,q_\mu\bigg\} \gamma_5
\Lambda_b(P) \,, \,\,  \label{Lambdab to proton: QCD}
\end{eqnarray}
with $q=P-P^{\prime}$. Following the Ref. \cite{Manohar:2000dt},
we rescale the hadronic states and spinors in the following way
\begin{eqnarray}
| H_Q(P) \rangle &=& \sqrt{m_Q} \, [ \, | H(v) \rangle +
O(1/m_{Q}) \, ]\,,
\nonumber \\
u_Q(P,\lambda) &=& \sqrt{m_Q} \, u(v,\lambda)\,,
\end{eqnarray}
so that  the heavy-quark mass dependence is removed from the
states and spinors labelled by the subscript $Q$.  Then, one can
immediately find that the QCD matrix element is equal to the one
in the effective theory up to a $O(1/m_{Q})$ correction. Comparing
two definitions (\ref{Lambdab to proton: EFT}) and (\ref{Lambdab
to proton: QCD}), one can find the following relations among the
form factors
\begin{eqnarray}
f_1(q^2)&=&g_1(q^2)= \eta(P^{\prime} \cdot v)\,, \nonumber \\
f_2(q^2)&=&g_2(q^2)=f_3(q^2)=g_3(q^2)=0\,. \label{relations:
Lambda_b to proton}
\end{eqnarray}
Numerical results of $\Lambda_b \to p$  form factors from
light-cone sum rules  (LCSR) \cite{Khodjamirian:2011jp}  have been
collected in Table \ref{tab_resLambdab} implying that the
relations derived from heavy-quark  and large-energy symmetries
are well respected \footnote{We stress that LCSR predictions of
the $\Lambda_b \to p$ form factors $f_1(0)$ and $g_1(0)$ presented
in \cite{Khodjamirian:2011jp} are in good agreement with that
extracted from  CDF measurement of non-leptonic decay $\Lambda_b
\to p \, \pi$ \cite{Aaltonen:2008hg} in the factorization limit. A
rather small value of the form factor $f_1 (0)=(2.3^{+0.6}_{-0.5})
\times 10^{-2}$ \cite{Wang:2009hra} from LCSR with $\Lambda_b$
distribution amplitudes is probably  due to the fact that the sum
rules are constructed from the correlation function involving  the
nucleon interpolating current $\eta=(u^{\rm T} C \not \! z )
\gamma_5 \not \! z d$ which couples to  both $\Delta$-resonances
and negative-parity baryons. For  more detailed discussions on
this issue, we refer the reader to  \cite{Khodjamirian:2011jp}.}.
\begin{table}[tb]
\begin{center}
\begin{tabular}{|c|c|c|}
  \hline
  \hline
   form factors &  $\eta_{\Lambda_b}^{(\mathcal{A})}$ & $\eta_{\Lambda_b}^{(\mathcal{P})}$  \\
   \hline
   &&\\
   $f_1(0)$ & $0.14^{+0.03}_{-0.03}$ &   $0.12^{+0.03}_{-0.04}$   \\
   && \\
   $f_2(0)$  & $-0.054^{+0.016}_{-0.013}$ &   $-0.047^{+0.015}_{-0.013}$ \\
   && \\
   \hline
   &&\\
   $g_1(0)$ &  $0.14^{+0.03}_{-0.03}$ &  $0.12^{+0.03}_{-0.03}$  \\
&&\\
   $g_2(0)$  & $-0.028^{+0.012}_{-0.009}$ & $-0.016^{+0.007}_{-0.005}$ \\
&&\\
  \hline
  \hline
\end{tabular}
\end{center}
\caption{Numerical results of $\Lambda_b \to p$  transition form
factors at zero momentum transfer calculated  in LCSR with
different interpolating currents of $\Lambda_b$ baryon from
\cite{Khodjamirian:2011jp}.} \label{tab_resLambdab}
\end{table}
It needs to stress that two form factors are needed to
parameterize  $\Lambda_b \to p$ transition matrix elements if only
heavy-quark spin symmetry is employed. Large-energy symmetry can
further simplify the long-distance physics of $\Lambda_b \to p$
transitions in the large-recoil region.

As mentioned  above, the relations shown in Eq. (\ref{relations:
Lambda_b to proton}) are only valid for soft form factors at
leading-order of $\Lambda_{\rm QCD}/m_{Q}$ and $\alpha_s$, and
they can be violated taking into account  the hard interactions. 
In other words, these nontrivial relations hold only for the
Feynman-mechanism contribution to the form factors, which are
suppressed by the nucleon distribution amplitudes in the end-point
region. In view of this observation, it is also possible to derive
the scaling behavior of the soft form factor $\eta(P^{\prime}
\cdot v)$ in the heavy quark limit making use of the asymptotic
behavior of nucleon distribution amplitudes. To the lowest
conformal spin,  the distribution amplitudes of nucleon are given
by \cite{Braun:2000kw}
\begin{eqnarray}
V_{N}(x_1, x_2, x_3)=T_{N}(x_1, x_2, x_3) = \phi^{asy}(x_1, x_2,
x_3) \equiv 120 x_1 x_2 x_3 \,.
\end{eqnarray}
Integrating over the end-point region for the  momentum fraction
of the recoiled quark $x_1 \sim 1- \Lambda / E $, one can obtain
the scaling behavior
\begin{eqnarray}
\eta(P^{\prime} \cdot v) \sim \int_{1- {\Lambda \over E}}^{1} d
x_1 \int_0^{1-x_1} d x_2 \, \phi^{asy}(x_1, x_2, 1-x_1-x_2) \sim
{\Lambda^3 \over E^3} \,. \label{scaling law: lambda_b to proton}
\end{eqnarray}

Now, we  will show that the scaling law of $\eta(P^{\prime} \cdot
v) $ derived above is compatibles with that from the LCSR. Very
recently,  $\Lambda_b \to p$ transition form factors were
revisited in Ref. \cite{Khodjamirian:2011jp} from the improved sum
rules approach, where the contribution of ground state
negative-parity bottom baryon has been separated out from the
$\Lambda_b$ contribution without absorbing it into the continuum
in the hadronic dispersion relation. An advantage of this approach
is that the physical form factors are insensitive to the specific
choice of the interpolating current of the heavy baryon as
observed from Table \ref{tab_resLambdab}.

To work out  $\Lambda_b \to p$ form factors at $q^2=0$ in the
heavy quark limit $m_b \to \infty$, we rescale Borel mass,
threshold parameter and the decay constant of heavy baryon
$\lambda_{\Lambda_b}^{(i)}$ following the Ref. \cite{Bagan:1997bp}
\begin{eqnarray}
M^2= 2 m_b \tau \,, & \qquad & s_0=m_b^2 + 2 m_b \omega_0 \,,
\nonumber \\
\lambda_{\Lambda_b}^{(i)} = \frac{\tilde{f}_{\Lambda_b}}{m_b}\,, &
\qquad  & m_{\Lambda_b}-m_b = \bar{\Lambda} \,.
\end{eqnarray}
Here, $\tau$ and $\omega_0$ correspond to the nonrelativistic
Borel mass and threshold parameter. The decay constants
$\lambda_{\Lambda_b}^{(i)}$ and $\tilde{f}_{\Lambda_b}$ are given
by
\begin{eqnarray}
 \langle 0 | \eta^{(i)}_{\Lambda_b} | \Lambda_b (P) \rangle =
m_{\Lambda_b}\lambda_{\Lambda_b}^{(i)}\,\, \Lambda_b(P)\,, \qquad
 \langle 0 | (u \, C \, \Gamma \, d ) \, b_v| \Lambda_b (v)
\rangle = \tilde{f}_{\Lambda_b} \Lambda_b (v) \,.
\end{eqnarray}
It is clear that $\Lambda_b$ decay constants defined by various
currents degenerate  in the heavy-quark limit. We then derive the
sum rules of the form factor $f_1(0)$ in the heavy quark limit
\begin{eqnarray}
f_1(0)=  \frac{12 }{m_b^3 \tilde{f}_{\Lambda_b}}  \left\{
\begin{array}{l}
 (\lambda_2 - 2 \lambda_1 ) \, m_N \, e^{\bar{\Lambda}/\tau} \,
\int_0^{\omega_0} d \omega \, \omega^2
e^{-\omega/\tau} \,\,\,  \,, \\
\\ 20 f_N \,
e^{\bar{\Lambda}/\tau}\, \int_0^{\omega_0} d \omega \, \omega^3
e^{-\omega/\tau} \,\,\,    \,,
\end{array}\right. \,
\label{LCSR: heavy quark limit}
\end{eqnarray}
where the upper (lower)  sum rule is constructed from the
correlation function with pseudoscalar (axial-vector) $\Lambda_b$
current  and the nucleon decay constants  $f_N$, $\lambda_1$ and
$\lambda_2$ are defined as \cite{Braun:2000kw}
\begin{eqnarray}
\langle 0| \epsilon^{ijk} \, [u^i(0) C \not \! \bar{n} u^j(0)] \,
\gamma_5  \not \! \bar{n} \, d_\gamma^k(0) | N(P)\rangle &=& f_N
(\bar{n} \cdot P) \not \! \bar{n} N(P) \,, \nonumber  \\
\langle 0| \epsilon^{ijk} \, [u^i(0) C \gamma_{\mu} u^j(0)] \,
\gamma_5 \gamma^{\mu} \, d_\gamma^k(0) | N(P)\rangle &=& \lambda_1
m_N N(P)\,,  \nonumber \\
\langle 0| \epsilon^{ijk} \, [u^i(0) C \sigma_{\mu \nu} u^j(0)] \,
\gamma_5 \sigma^{\mu \nu} \, d_\gamma^k(0) | N(P)\rangle &=&
\lambda_2 m_N N(P) \,.  \nonumber
\end{eqnarray}
Comparing Eq. (\ref{scaling law: lambda_b to proton}) with Eq.
(\ref{LCSR: heavy quark limit}), one can observe that LCSR and
HQET/SCET formalism predict consistent scaling behavior of the
soft baryonic form factor $\eta(P^{\prime} \cdot v) \sim
(\Lambda_{\rm QCD}/m_Q)^3$, which is different from the scaling of
heavy-to-light mesonic soft form factor $\xi(P^{\prime} \cdot v)
\sim (\Lambda_{\rm QCD}/m_Q)^{3/2}$ following from the symmetry
argument \cite{Charles:1998dr}. Assuming the scaling of the inner
sum rule parameters $\omega_0 \sim \tau \sim  m_N$, we can extract
the relation of nucleon decay constants $f_N$, $\lambda_1$ and
$\lambda_2$
\begin{eqnarray}
{ \lambda_2 - 2 \lambda_1  \over  f_N} = 14.2 \,,
\end{eqnarray}
from the matching of two sum rules presented in Eq. (\ref{LCSR:
heavy quark limit}), which is consistent with the prediction from
two-point QCD sum rules \cite{Braun:2000kw}.

Applying the same technique to $\Lambda_b \to \Lambda$ form
factors, we obtain
\begin{eqnarray}
\langle \Lambda(P^{\prime}) | \bar{s} \Gamma b| \Lambda_b(v))
\rangle = \bar{\Lambda}(P^{\prime}) \Gamma \Lambda_b(v) \, {\rm
Tr} [ \not \! n \gamma_5 \mathcal{\bar{J}} ]=
\bar{\eta}(P^{\prime} \cdot v) \bar{\Lambda}(P^{\prime}) \Gamma
\Lambda_b(v) \,, \,\,\,\, \label{Lambdab to Lambda}
\end{eqnarray}
indicating that ten $\Lambda_b \to \Lambda$ form factors can be
reduced to one universal form factor in the large recoil region in
the heavy-quark limit. This can be also easily understood from the
fact that the spin of the strange quark is the same as that of the
$\Lambda$ baryon in the quark model and the effective Lagrangian
describing the interaction of energetic quarks with soft gluon
does not involve nontrivial Dirac dynamics. Similar observation
was also made in \cite{Hiller:2001zj} with slightly different
arguments.

For $\Lambda_b \to \Sigma$ transition, the hadronic matrix element
 in the heavy quark limit can be simplified as
\begin{eqnarray}
\langle \Sigma(P^{\prime}) | \bar{s} \Gamma b| \Lambda_b(v))
\rangle &=& \bar{\Sigma}(P^{\prime}) \gamma_5 \Gamma \Lambda_b(v)
\, {\rm Tr} [  \not \! n  \mathcal{\tilde{J}}_1 ]+
\bar{\Sigma}(P^{\prime}) \gamma^{\mu} \gamma_5 \Gamma \Lambda_b(v)
\, {\rm Tr} [ n^{\nu} i \sigma_{\mu \nu} \mathcal{\tilde{J}}_2 ]
\nonumber \\
&=& 0 \,, \label{Lambdab to Sigma}
\end{eqnarray}
as a consequence of the space-time parity symmetry, stating that
all the ten $\Lambda_b \to \Sigma$ transition form factors
should vanish in the large recoil limit. 

\subsection{$\Omega_b \to \Xi$ and $\Omega_b \to \Omega$  form factors}

Repeating the same procedure for  $\Omega_b \to \Xi$ transition,
we can  obtain
\begin{eqnarray}
\langle  \Xi(P^{\prime}) | \bar{u} \Gamma b| \Omega_b(v) \rangle
&=& \bar{\Xi}(P^{\prime}) \gamma_5 \Gamma  R^{\rho}_{1/2}(v) \,
{\rm
Tr} [ \not \! n (\mathcal{K}_1)_{\rho}]  \nonumber \\
&& + \bar{\Xi}(P^{\prime}) \gamma^{\mu} \gamma_5 \Gamma
 R^{\rho}_{1/2}(v)  \, {\rm Tr} [ i  n^{\nu} \sigma_{\mu \nu}
(\mathcal{K}_2)_{\rho} ] \,,
 \label{Omega_b to Xi}
\end{eqnarray}
where again the nonperturbative functions $\mathcal{K}_i$
($i=1,2$)  involve the matrices $\not \! n$ and $\not \! v$,
however they are independent of the Dirac structure $\Gamma$ of
the transition current. Using the equation of motion $\not \! n
\xi_n =0$ and performing the replacement $R_{1/2}^{\mu} \to {1
\over \sqrt{3}} (\gamma^{\mu}+v^{\mu}) \gamma_5 h_v$, the above
hadronic matrix element can be simplified as
\begin{eqnarray}
\langle  \Xi(P^{\prime}) | \bar{u} \Gamma b| \Omega_b(v) \rangle
&=& \zeta_1(P^{\prime} \cdot v) \, \bar{\Xi}(P^{\prime}) \gamma_5
\Gamma {n_{\mu} \over \sqrt{2}} (\gamma^{\mu}+v^{\mu}) \gamma_5
\Omega_b(v)
\nonumber \\
&& + \zeta_2(P^{\prime} \cdot v) \, \bar{\Xi}(P^{\prime}) \gamma_5
\not \! \bar{n} \Gamma {n_{\mu} \over \sqrt{2}}
(\gamma^{\mu}+v^{\mu}) \gamma_5 \Omega_b(v)
\nonumber \\
&& + \zeta_3(P^{\prime} \cdot v) \, \bar{\Xi}(P^{\prime}) \gamma_5
\gamma^{\mu} \Gamma  (\gamma^{\mu}+v^{\mu}) \gamma_5 \Omega_b(v)
 \label{Omega_b to Xi: EFT} \,,
\end{eqnarray}
in the large recoil limit.

Similarly, one can derive the $\Omega_b \to \Omega$ transition
matrix element  in the heavy quark limit as
\begin{eqnarray}
\langle  \Omega(P^{\prime}) | \bar{s} \Gamma b| \Omega_b(v)
\rangle &=& \bar{\Omega}^{\mu} (P^{\prime})  \Gamma
R^{\rho}_{1/2}(v)  \, {\rm Tr} [ \gamma_{\mu}
(\mathcal{\bar{K}}_1)_{\rho} ] + \bar{\Omega}^{\nu}(P^{\prime})
\gamma^{\mu}  R^{\rho}_{1/2}(v) \, {\rm Tr} [ i  \sigma_{\mu \nu}
(\mathcal{\bar{K}}_2)_{\rho} ]
\nonumber \\
&& + \bar{\Omega}^{\nu}(P^{\prime}) n^{\mu}  \Gamma
R^{\rho}_{1/2}(v) \, {\rm Tr} [ i   \sigma_{\mu \nu}
(\mathcal{\bar{K}}_3)_{\rho} ] \,.
 \label{Omega_b to Omega}
\end{eqnarray}
Making use of the equation of motion, we obtain
\begin{eqnarray}
\langle  \Omega(P^{\prime}) | \bar{s} \Gamma b| \Omega_b(v)
\rangle &=& \bar{\zeta}_1(P^{\prime} \cdot v) \,
\bar{\Omega}^{\mu}(P^{\prime}) \, \bar{n}_{\mu}  \Gamma n_{\rho}
\, (\gamma^{\rho}+v^{\rho}) \gamma_5 \Omega_b(v) \, \nonumber \\
&& + \bar{\zeta}_2(P^{\prime} \cdot v) \,
\bar{\Omega}^{\mu}(P^{\prime})  \,
g_{\mu \rho} \, \Gamma \, (\gamma^{\rho}+v^{\rho}) \gamma_5 \Omega_b(v)  \, \nonumber \\
&& + \bar{\zeta}_3(P^{\prime} \cdot v) \,
\bar{\Omega}^{\mu}(P^{\prime})  \, {\bar{n}_{\mu}  \over \sqrt{2}}
\not \! \bar{n}
\Gamma n_{\rho} \, (\gamma^{\rho}+v^{\rho}) \gamma_5 \Omega_b(v)  \, \nonumber \\
&& + \bar{\zeta}_4(P^{\prime} \cdot v) \,
\bar{\Omega}^{\mu}(P^{\prime}) \, {\bar{n}_{\mu} \over \sqrt{2} }
\gamma_{\rho}
\Gamma \, (\gamma^{\rho}+v^{\rho}) \gamma_5 \Omega_b(v)  \, \nonumber \\
&& + \bar{\zeta}_5(P^{\prime} \cdot v) \,
\bar{\Omega}^{\mu}(P^{\prime})  \, g_{\mu \rho} {\not \! \bar{n}
\over \sqrt{2}} \, \Gamma \, (\gamma^{\rho}+v^{\rho}) \gamma_5
\Omega_b(v) \,,
 \label{Omega_b to Omega: EFT}
\end{eqnarray}
where five soft form factors are necessary to parameterize the
nonperturbative dynamics, and the normalization factor
``$\,1/\sqrt{2}\,$" is introduced for later convenience. Without
employing the large-energy symmetry, an additional soft form
factor associating with the spin structure
\begin{eqnarray}
\bar{\Omega}^{\mu}(P^{\prime}) \, \bar{n}_{\mu}  \,
\gamma_{\rho}\, \not \! v  \,  \Gamma  \, (\gamma^{\rho}+v^{\rho})
\gamma_5 \Omega_b(v)
\end{eqnarray}
should be included as shown  in \cite{Hussain:1992rb}.

\section{Applications to FCNC $\Lambda_b$ and $\Omega_b$ decays}

\subsection{Rare decays of  $\Lambda_b \to
\Lambda \, \gamma$ and $\Lambda_b \to \Lambda \, l^{+} l^{-}$ }

\subsubsection{Radiative decay $\Lambda_b \to \Lambda \, \gamma$}

The underlying flavour-changing $b\to s$ transition in the
Standard Model  (SM) is described by the effective Hamiltonian:
\begin{eqnarray}
 H_{eff}=
-\frac{4G_{F}}{\sqrt{2}}V_{tb}V_{ts}^{\ast}
{\sum\limits_{i=1}^{10}} C_{i}({\mu}) O_{i}({\mu})\,,
\label{eq:Heff}
\end{eqnarray}
a linear combination of the effective operators $O_i$ weighted by
their Wilson coefficients $C_i$. The leading contribution of
$\Lambda_b \to \Lambda \gamma$ is generated by the electromagnetic
penguin operator
\begin{eqnarray}
O_{7\gamma}&=& -{e \over 16 \pi^2} \bar{s} \sigma_{\mu \nu} (m_s
L+ m_b R)b F^{\mu \nu }\,,
\end{eqnarray}
with the notation $L(R)=\frac{1-(+)\gamma_5}{2}$. Allowing for
couplings beyond the SM, we introduce a more general dipole
transition operator
\begin{eqnarray}
\tilde{O}_{7\gamma}&=& -{e \over 32 \pi^2} m_b \, \bar{s}
\sigma_{\mu \nu} (g_V- g_A \gamma_5)b F^{\mu \nu }\,.
\end{eqnarray}

Considering  the  $\Lambda$-baryon polarization asymmetry   in
$\Lambda_b \to \Lambda + \gamma$, we firstly define   the
four-spin vector $s^{\mu}$ of $\Lambda$ baryon in its rest frame
\begin{eqnarray}
(s^{\mu})_{r.s.}=(0, \,\, \hat{\bf{\xi}}), \label{spin vector of
lambda baryon 1}
\end{eqnarray}
which can be  boosted into the rest frame of $\Lambda_b$ baryon
\begin{eqnarray}
s^{\mu}=({{\bf{P}}_{\Lambda}\cdot {\hat{\bf{\xi}}} \over
m_{\Lambda}}, \,\, {\bf{\hat{\xi}}}+ {s_0 \over
E_{\Lambda}+m_{\Lambda}} {\bf{P}}_{\Lambda}), \label{spin vector
of lambda baryon 2}
\end{eqnarray}
with  ${\bf{P}}_{\Lambda}$ and $E_{\Lambda}$ being the
three-momentum and energy of $\Lambda$ baryon. It is
straightforward  to derive that
\begin{eqnarray}
v \cdot s= {1-x_{\Lambda}^2 \over 1+x_{\Lambda}^2} {\bf{\hat{p}}}
\cdot {\bf{s}} = {1-x_{\Lambda}^2 \over 2x_{\Lambda}}
{\bf{\hat{p}}} \cdot {\bf{\hat{\xi}}}, \label{spin relation}
\end{eqnarray}
where $x_{\Lambda}=m_{\Lambda}/m_{\Lambda_b}$ and ${\bf{\hat{p}}}$
is a unite vector along the direction of $\Lambda$-baryon
momentum.

Following Refs. \cite{Mannel:1997xy,Huang:1998ek}, the polarized
decay width of $\Lambda_b \to \Lambda + \gamma$ has a form
\begin{eqnarray}
\Gamma(\Lambda_b \to \Lambda \gamma) = {1 \over 2}\Gamma_0
[1+\alpha \,\, {\bf{\hat{p}}} \cdot {\bf{s}} ]= {1 \over
2}\Gamma_0 [1+\alpha^{\prime} \,\, {\bf{\hat{p}}} \cdot
{\bf{\hat{\xi}}} ], \label{form of polarization asymmetry}
\end{eqnarray}
where $\Gamma_0$  is the total decay with of $\Lambda_b \to
\Lambda + \gamma$
\begin{eqnarray}
\Gamma_0= {  G_F^2 \alpha_{em}\over 64 \pi^4 }
|V_{tb}V_{ts}^{\ast}|^2 m_b^2 m_{\Lambda_b}^3 (1-x^2_{\Lambda})^3
\big|\bar{\eta}(P^{\prime} \cdot v)\big|^2
(|g_V|^2+|g_A|^2)|\tilde{C}^{eff}_{7 \gamma}|^2
\end{eqnarray}
and the polarization asymmetry  reads
\begin{eqnarray}
\alpha={2x_{\Lambda} \over
1+x_{\Lambda}^2}\alpha^{\prime}={2x_{\Lambda} \over
1+x_{\Lambda}^2}{2 g_V g_A \over g_V^2+g_A^2 } + O\big(
{\Lambda_{\rm QCD}\over m_b} \big) + O\big( \alpha_s \big) \,.
\label{polarization asymmetry of radiative decay}
\end{eqnarray}
Taking the soft form factor $\bar{\eta}(q^2=0)
=0.15^{+0.02}_{-0.02}$ from QCD LCSR \cite{Wang:2008sm} and
neglecting the long-distance contribution,\footnote{A preliminary
study on  some possible long-distance contributions, for instance
charm-quark loop, internal $W$-exchange and light-quark loop, was
already performed in \cite{Mannel:1997xy}, where long-distance
contribution was found to be suppressed either by the large
virtuality of the charm propagators or by the CKM matrix
elements.} we predict the branching ratio  ${\rm BR} (\Lambda_b
\to \Lambda + \gamma)=\big(7.7^{+2.2}_{-1.9}\big) \times 10^{-6}$
in the SM, which is compatible with that estimated in
\cite{Mannel:1997xy} based on the heavy-to-light baryonic form
factors from the pole model. We also mention in passing that
$\Lambda$-baryon polarization asymmetry, to the leading order, is
only determined by the short distance coefficients of partonic
transition as already observed in \cite{Huang:1998ek,Chua:1998dx}.

\subsubsection{Semileptonic  decay $\Lambda_b \to \Lambda \, l^{+} l^{-}$}

The dominant contributions to $\Lambda_b \to \Lambda \, l^{+}
l^{-}$ are generated by the operators $O_{7\gamma}$ and $O_{9,10}$
\begin{eqnarray}
O_9 = \frac{\alpha_{em}}{4\pi}\left(\bar{s} \gamma_\rho L b
\right) \left( \bar{l}\gamma^{\rho}l \right)\,, \qquad
O_{10}=\frac{\alpha_{em}}{4\pi}\left(\bar{s} \gamma_\rho L b
\right) \left(\bar{l}\gamma^{\rho}\gamma_5 l \right)\,.
\end{eqnarray}
The free quark decay amplitude for $b \to s l^{+}l^{-}$ process
reads
\begin{eqnarray}
\mathcal{A}(b\to s l^+l^-) &=&
\frac{G_{F}}{\sqrt{2}}V_{tb}V_{ts}^{*} {\alpha_{em}\over \pi}\bigg
\{-{2i \over q^2}C_7^{eff}(\mu) \bar{s} \sigma_{\mu\nu}q^{\nu}(m_b
R+m_sL) b  \, \bar{l}\gamma^{\mu}l \nonumber
\\&&+ C_{9}^{eff}(\mu)\bar{s}\gamma_{\mu}Lb \,
\bar{l}\gamma^{\mu}l+C_{10} \bar{s}\gamma_{\mu}Lb \,
\bar{l}\gamma^{\mu}\gamma_5l \bigg\}. \label{b to s l l}
\end{eqnarray}
In  the leading-order, the hadronic decay amplitude
${\mathcal{A}}_{\Lambda_b \to \Lambda l^+ l^-}$   can be  derived
by sandwiching the free quark amplitude (\ref{b to s l l}) between
the initial and final states. Defining the  differential
 forward-backward asymmetry  for the semileptonic decay
\begin{eqnarray}
{d A_{FB}(q^2) \over d q^2}=\int_0^1 dz {d^2 \Gamma (q^2, z) \over
dq^2 dz} - \int_{-1}^0 dz  {d^2 \Gamma (q^2, z) \over dq^2 dz}\,,
\end{eqnarray}
we obtain the following expression for $\Lambda_b \to \Lambda l^+
l^-$ transition in the SM
\begin{eqnarray}
{d A_{FB}(\Lambda_b \to \Lambda + l^{+} l^{-}) \over d q^2}={
G_F^2 \alpha_{em}^2 \over 256 m_{\Lambda_b}^3 \pi^5} |V_{tb}
V_{ts}^{\ast}|^2
 \lambda(m_{\Lambda_b^2},m_{\Lambda}^2,q^2) (1-{4 m_l^2 \over q^2})
R_{FB}(q^2),
\end{eqnarray}
with $\lambda(a,b,c)=a^2+b^2+c^2-2ab-2ac-2bc$ and
\begin{eqnarray}
 R_{FB}(q^2)  &=& \big [2 m_b \, m_{\Lambda_b}  {\rm
{Re}}(C_{7 \gamma}^{eff} C_{10}^{\ast}) + q^2 {\rm {Re}}(C_9^{eff}
C_{10}^{\ast}) \big ] \big|\bar{\eta}(P^{\prime} \cdot v)\big|^2  \nonumber \\
&& + O\big( {\Lambda_{\rm QCD}\over m_b} \big) + O\big( \alpha_s
\big). \label{FBA expressions}
\end{eqnarray}
It is obvious that the differential forward-backward asymmetry in
$\Lambda_b \to \Lambda l^{+} l^{-}$ decay only depends on the
Wilson coefficients ${\rm {Re}}(C_7^{eff} C_{10}^{\ast})$ and
${\rm {Re}}(C_9^{eff} C_{10}^{\ast})$. The zero-position  $t_0$ of
forward-backward asymmetry is given  by
\begin{eqnarray}
t_0 (\Lambda_b \to \Lambda + l^{+} l^{-}) = - 2 \, m_b\,
m_{\Lambda_b} \, \frac {{\rm {Re}}(C_{7\gamma}^{eff}
C_{10}^{\ast})} {{\rm {Re}}(C_9^{eff} C_{10}^{\ast})} + O\big(
{\Lambda_{\rm QCD}\over m_b} \big) + O\big( \alpha_s \big)\,.
\label{AFB:Lambda_b}
\end{eqnarray}
The expression of $t_0 (\Lambda_b \to \Lambda + l^{+} l^{-})$ is
exactly the same as that in the case of $B \to K^{\ast} l^{+}
l^{-}$\cite{Ali:1999mm}, and it is free of hadronic uncertainties
in the large recoil limit of $\Lambda$ baryon. Substituting the
values of Wilson coefficients at next-to-leading-logarithmic order
as well as the pole-mass of the $b$ quark $m_b$
\cite{Beneke:2001at}, we find $t_0 (\Lambda_b \to \Lambda + l^{+}
l^{-})=3.8 \,\, {\rm GeV^2}$ which is consistent with those
derived in \cite{Wang:2008sm} using $\Lambda_b \to \Lambda$ form
factors from LCSR.

\subsection{Rare decays of  $\Omega_b \to \Omega \, \gamma$ and $\Omega_b \to \Omega \, l^{+} l^{-}$ }

\subsubsection{Radiative decay $\Omega_b \to \Omega \, \gamma$}

Taking into account the left-hand coupling, the decay amplitude of
$\Omega_b \to \Omega \, \gamma$ can be written as
\begin{eqnarray}
\mathcal{A}(\Omega_b \to \Omega \,
\gamma)=\frac{4G_{F}}{\sqrt{2}}V_{tb}V_{ts}^{\ast} {e \over 32
\pi^2} m_b \langle \Omega| \tilde{C}^{eff}_{7 \gamma} \,
\tilde{O}_{7\gamma} | \Omega_b \rangle \,.
\end{eqnarray}
In terms of the hadronic matrix element given in Eq. (\ref{Omega_b
to Omega: EFT}),  one can derive the decay width of the radiative
decay $\Omega_b \to \Omega \, \gamma$ as
\begin{eqnarray}
\Gamma(\Omega_b \to \Omega \, \gamma) &=&{ G_F^2 \alpha_{em} \over
384 \pi^4} |V_{tb}V_{ts}^{\ast}|^2 m_b^2 m_{\Omega_b}^3
{(1-x_{\Omega}^2)^3 \over x_{\Omega}^2 } \big \{ \bar{\zeta}_1^2 +
2 \bar{\zeta}_1 \big [ (1+x_{\Omega}^2)  \bar{\zeta}_2 -
x_{\Omega}\bar{\zeta}_4 \big ] \nonumber \\
&& +(1+6 x_{\Omega}^2 + x_{\Omega}^4)\bar{\zeta}_2^2 +
x_{\Omega}^2 \bar{\zeta}_2^4 - 2 x_{\Omega} (1+3x_{\Omega}^2)
\bar{\zeta}_2 \bar{\zeta}_4 \big \} \, \nonumber \\
&& \times (|g_V|^2+|g_A|^2)|\tilde{C}^{eff}_{7 \gamma}|^2\,,
\end{eqnarray}
where $x_{\Omega}=m_{\Omega}/m_{\Omega_b}$ and the polarization
sum of Rarita-Schwinger spin vectors
\begin{eqnarray}
\Omega_{\mu}(P^{\prime}) \bar{\Omega}_{\nu}(P^{\prime})= - (\not
\! P^{\prime} + m_{\Omega})\, \bigg [ g_{\mu \nu} - {1 \over 3}
\gamma_{\mu} \gamma_{\nu} -{2 \over 3 m_{\Omega}^2}
P_{\mu}^{\prime} P_{\nu}^{\prime}  + {
\gamma_{\nu}P_{\mu}^{\prime}-\gamma_{\mu}P_{\nu}^{\prime}\over 3
m_{\Omega}}\bigg ] \, .
\end{eqnarray}
has been employed. In the massless limit of $\Omega$ baryon, the
decay width is further simplified as
\begin{eqnarray}
\Gamma(\Omega_b \to \Omega \, \gamma) &=&{ G_F^2 \alpha_{em} \over
384 \pi^4} |V_{tb}V_{ts}^{\ast}|^2  {m_b^2 m_{\Omega_b}^3 \over
x_{\Omega}^2 } \big ( \bar{\zeta}_1^2 +\bar{\zeta}_2^2 \big )
(|g_V|^2+|g_A|^2)|\tilde{C}^{eff}_{7 \gamma}|^2\,,
\end{eqnarray}
indicating  that the radiative decay $\Omega_b \to \Omega \,
\gamma$ is strongly enhanced by  a  factor ${1 / x_{\Omega}^2}$
coming from the helicity-$1/2$ $\Omega$ contribution. Taking the
ratio of decay width between $\Omega_b \to \Omega \, \gamma$ and
$\Lambda_b \to \Lambda \, \gamma$ in the massless limit of light
baryon, we obtain
\begin{eqnarray}
\frac{\Gamma(\Omega_b \to \Omega \, \gamma)} {\Gamma(\Lambda_b \to
\Lambda \, \gamma)} &=& \frac{\bar{\zeta}_1^2 +\bar{\zeta}_2^2}{6
\, \bar{\eta}^2 } \cdot \big(\frac{m_{\Omega_b}}{m_{\Lambda_b}}
\big)^3  \cdot \big(\frac{m_{\Omega_b}}{m_{\Omega}} \big)^2 \,,
\end{eqnarray}
showing that radiative decay $\Omega_b \to \Omega \, \gamma$ is
probably the most promising FCNC $ b \to s$ radiative baryonic
decay channel and would be a golden channel to extract the
helicity structures of weak effective Hamiltonian.

\subsubsection{Semileptonic  decay $\Omega_b \to \Omega \, l^{+} l^{-}$}

The decay amplitude ${\mathcal{A}}_{\Omega_b \to \Omega l^+ l^-}$
responsible for   $\Omega_b \to \Omega \, l^{+} l^{-}$ transition
can be calculated following a similar way  for
${\mathcal{A}}_{\Lambda_b \to \Lambda l^+ l^-}$ albeit with more
involved spin structures. The differential forward-backward
asymmetry in  $\Omega_b \to \Omega \, l^{+} l^{-}$ decay is
calculated as
\begin{eqnarray}
{d A_{FB}(\Omega_b \to \Omega \, l^{+} l^{-}) \over d q^2}={ G_F^2
\alpha_{em}^2 \over 1536 m_{\Omega_b}^3 \pi^5} |V_{tb}
V_{ts}^{\ast}|^2
 \lambda^{1/2}(m_{\Omega_b^2},m_{\Omega}^2,q^2) \sqrt{1-{4 m_l^2 \over q^2}}
\tilde{R}_{FB}(q^2), \,\,\,\,\,
\end{eqnarray}
with
\begin{eqnarray}
 \tilde{R}_{FB}(q^2)  &=& -m_b \, m_{\Omega_b}^3 \, \bigg(1-{q^2 \over m_{\Omega_b}^2 }\bigg)^3
 \bigg(\frac{m_{\Omega_b}}{m_{\Omega}} \bigg)^2 \bigg \{ 2 \big [ \big ( \bar{\zeta}_1 +\bar{\zeta}_2 \big )^2
 - {q^2 \over m_{\Omega_b}^2 } \big( \bar{\zeta}_3+\bar{\zeta}_4+\bar{\zeta}_5 \big)^2 \big ]  {\rm
{Re}}(C_{7 \gamma}^{eff} C_{10}^{\ast}) \nonumber \\
&& + {q^2 \over m_b \, m_{\Omega_b} } \big [ \big ( \bar{\zeta}_1
+\bar{\zeta}_2 \big )^2 -  \big(
\bar{\zeta}_3+\bar{\zeta}_4+\bar{\zeta}_5 \big)^2 \big ]{\rm
{Re}}(C_9^{eff}
C_{10}^{\ast}) \bigg \}   \nonumber \\
&& + O\big( {\Lambda_{\rm QCD}\over m_b} \big) + O\big( \alpha_s
\big). \label{FBA expressions}
\end{eqnarray}
An estimate from the quark model \cite{Singer:1996xh} indicates
that the form factors $\bar{\zeta}_i$ ($i=3, \, 4, \, 5\,$) are
negligible numerically. In this case, one can easily derive the
zero-point of the forward-backward asymmetry
\begin{eqnarray}
t_0 (\Omega_b \to \Omega \, l^{+} l^{-})= - 2 \, m_b \,
m_{\Omega_b} \, \frac {{\rm {Re}}(C_7^{eff} C_{10}^{\ast})} {{\rm
{Re}}(C_9^{eff} C_{10}^{\ast})} + O\big( {\Lambda_{\rm QCD}\over
m_b} \big) + O\big( \alpha_s \big)\,,
\end{eqnarray}
which is again free of hadronic uncertainties in  leading power of
${\rm \Lambda}_{\rm QCD}/m_b$  and in leading order of $\alpha_s$.
It is manifest that the expression of $t_0$ in $\Omega_b \to
\Omega \, l^{+} l^{-}$ is the same as the one for semileptonic
$\Lambda_b \to \Lambda \, l^{+} l^{-}$ decay shown in Eq.
(\ref{AFB:Lambda_b}). Substituting the values of the Wilson
coefficients yields $t_0 (\Omega_b \to \Omega \, l^{+} l^{-})=4.1
\,\, {\rm GeV^2}$ numerically, where the uncertainties are not
expected to be larger than ${\rm \Lambda}_{\rm QCD}/m_b$ owing to
the cancellation of nonperturbative effect in the forward-backward
asymmetry. More dedicated work on $\Omega_b \to \Omega$ transition
form factors from nonperturbative approaches based on QCD, such as
Lattice QCD and QCD LCSR, is highly demanded to provide nontrivial
tests of the predictions presented here.

\section{Discussion}

Weak decays of heavy baryons containing a bottom quark are among
the topics of central interest in heavy flavor physics for many
reasons. In contrast to the $B$-meson decays, an attractive
peculiarity of these decays is that they allow the study of  spin
correlation providing an unique ground to extract the helicity
structure of the flavor changing currents. Heavy-to-light baryon
form factors embedding the  long-distance hadronic dynamics  are
essential to describe semileptonic bottom baryon decays and also
enter into the factorization formulae of nonleptonic bottom baryon
decays. On account of a large amount of baryonic  form  factors in
QCD, reduction of independent nonperturbative functions with the
help of an  effective theory can help to simplify complicated
infrared dynamics in specific kinematical limit. One  classical
example is that in the small recoil limit two soft form factors
are adequate to parameterize the hadronic matrix element
responsible for $\Lambda_b \to \Lambda$ transition  in  HQET
\cite{Mannel:1990vg}. It was the aim of this work to explore the
relations among heavy-to-light baryonic form factors in the
opposite kinematical limit, the large recoil region.

We  discuss the tensor representations for   bottom baryons in the
heavy-quark limit following \cite{Mannel:1990vg,Falk:1991nq} and
then work out the light-cone projectors for both baryon-octet and
baryon-decuplet. With the tensor formalism introduced in
\cite{Falk:1990yz}, we show that only one form factor is essential
to parameterize  $\Lambda_b \to p$ and $\Lambda_b \to \Lambda$
matrix elements in the heavy quark limit and in the large energy
limit of the  light baryon; while $\Lambda_b \to \Sigma$
transition form factors should vanish in the same limit due to the
violation of space-time parity symmetry. The scaling behavior of
the $\Lambda_b \to p$ soft form factor is also derived with the
nucleon distribution amplitudes in the leading conformal spin and
the yielding scaling law $\xi(P^{\prime} \cdot v) \sim
(\Lambda_{\rm QCD}/m_Q)^{3}$ is exactly the same as that derived
from QCD LCSR. We then observe that three form factors are needed
to describe $\Omega_b \to \Xi$ decays and five form factors are
required to parameterize $\Omega_b \to \Omega$ transition matrix
element remembering that one additional form factor should be
included in the small recoil region.

Applying  the relations of form factors to  FCNC transitions
$\Lambda_b \to \Lambda \, \gamma$ and $\Lambda_b \to \Lambda \,
l^{+} l^{-}$,  we confirm that the polarization asymmetry  in
radiative $\Lambda_b \to \Lambda \, \gamma$ decay is free of
hadronic uncertainties in the leading power of $1/m_b$ and in the
leading order of  $\alpha_s$; the forward-backward asymmetry of
$\Lambda_b \to \Lambda \, l^{+} l^{-}$ is the same as the one in
$B \to K^{\ast} \, l^{+} l^{-}$ to the same accuracy. It is very
interesting to notice that the radiative decay $\Omega_b \to
\Omega \, \gamma$ is strongly enhanced by a  factor of
$m_{\Omega_b}^2/m_{\Omega}^2$ contributed  from  a helicity-$1/2$
$\Omega$ baryon and this channel would be among the most valuable
probes on the chirality information of short-distance dipole
transition. Forward-backward asymmetry of semileptonic $\Omega_b
\to \Omega \, l^{+} l^{-}$ decay is generally dependent on the
long-distance hadronic dynamics reflected in the transition form
factors, however, this asymmetry will be only determined by the
short-distance coefficients in an exact form of that for
$\Lambda_b \to \Lambda \, l^{+} l^{-}$ channel provided that the
form factors $\bar{\zeta}_i$ ($i=3, \, 4, \, 5\,$) are negligible
as indicated from the quark model.


\vspace{.3cm}

\noindent \textit{Note added:} While completing this work we have
been informed of related work by Thorsten~Feldmann and Matthew W\,
Y\, Yip \cite{Feldmann}, where they compute $\Lambda_b \to
\Lambda$ transition form factors in the framework of SCET sum
rules. These authors discuss similar issues and include also
symmetry breaking effects due to the sub-leading currents in SCET
for the relations of form factors discussed here.

\section*{Acknowledgement}

We are grateful to Matth\"{a}us Bartsch for helpful discussions
and to  Thorsten~Feldmann  for notifying us about their work prior
to publication. This work is supported by the German research
foundation DFG under contract MA1187/10- 1 and by the German
Ministry of Research (BMBF), contract 05H09PSF.

\end{document}